# Toward Whole-Brain Minimally-Invasive Vascular Imaging

Anatole Jimenez
*Physics for Medicine Paris, INSERM U1273, ESPCI, CNRS, PSL University*
Paris, France
anatole.jimenez@espci.fr

Bruno Osmanski
*Iconeus*
Paris, France
bruno.osmanski@iconeus.fr

Denis Vivien
*Physiopathology and Imaging of Neurological Disorders, GIP Cyceron*
Caen, France
vivien@cyceron.fr

Mickael Tanter
*Physics for Medicine Paris, INSERM U1273, ESPCI, CNRS, PSL University*
Paris, France
mickael.tanter@espci.fr

Thomas Gaberel
*Physiopathology and Imaging of Neurological Disorders, GIP Cyceron, Department of Neurosurgery, Université de Caen*
Caen, France
thomas.gaberel@hotmail.fr

Thomas Deffieux
*Physics for Medicine Paris, INSERM U1273, ESPCI, CNRS, PSL University*
Paris, France
thomas.deffieux@espci.fr

*Abstract*—Imaging the brain vasculature can be critical for cerebral perfusion monitoring in the context of neurocritical care. Although ultrasensitive Doppler (UD) can provide good sensitivity to cerebral blood volume (CBV) in a large field of view, it remains difficult to perform through the skull. In this work, we investigate how a minimally invasive burr hole, performed for intracranial pressure (ICP) monitoring, could be used to map the entire brain vascular tree. We explored the use of a small motorized phased array probe with a non-implantable preclinical prototype in pigs. The scan duration (18 min) and coverage (62 ± 12 % of the brain) obtained allowed global CBV variations detection (relative in brain $Doppler_{decrease}$ = -3[-4–+16]% & $Doppler_{increase}$ = +1[-3–+15]%, n = 6 & 5) and stroke detection (relative in core $Doppler_{stroke}$ = -25%, n = 1). This technology could one day be miniaturized to be implanted for brain perfusion monitoring in neurocritical care.

## I. Introduction

Traumatic brain injuries and aneurysm-related hemorrhages are significant causes of death and severe disabilities. Brain lesions significantly arise from secondary ischemic damage in comatose patients [1] [2]. The maintenance of the cerebral blood supply of those patients is the cornerstone of neurocritical care, and a reliable and early ischemia detection method is critical for appropriate intervention. Unfortunately, existing clinical modalities do not fulfill this task [3]. Although ultrasensitive Doppler (UD) can provide good sensitivity to cerebral blood volume (CBV) [4] [5], it remains challenging to perform through the skull. In this work, we investigate how a burr hole needed for intracranial pressure (ICP) monitoring [6] could be used to monitor whole brain vascularization using a preclinical prototype.

## II. Methods

### A. The preclinical prototype

We developed a preclinical prototype that consists of a phased array probe mounted on a motorized rotary unit, the motor's rotation axis aligned on the probe's depth axis (Fig 1a). The probe is a 64-element linear phased array with a 10-mm diameter circular aperture (5 MHz central frequency, 0.145 mm pitch, 9.1 mm aperture ($L$)). This prototype is not implantable. The probe was driven by an ultrasound research

Funded by AAPG2020 BrainCam Agence Nationale de Recherche

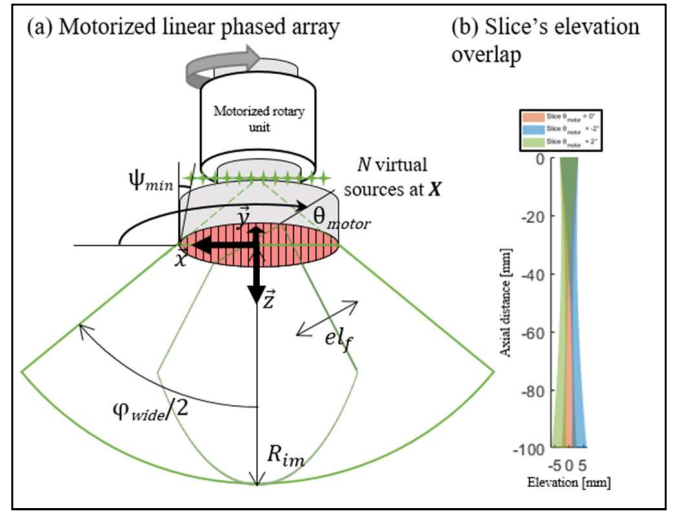

Fig 1. The preclinical prototype: (a) schematic view of the linear phased array mounted on a motorized rotary unit with imaging field of view of the slice at azimuth $\theta = 0°$ and $\theta = \theta_{motor}$, (b) consecutive slices in elevation overlapping.

system (Vantage 128, Verasonics, USA). Raw radio frequency (RF) data were saved and processed offline.

We performed 3D Doppler scans by acquiring UD slices at different rotation angles. An UD image was acquired with full aperture using $N$-diverging waves corresponding to different virtual sources at position $X = (x, z)$ set horizontally [7] at minimum $\psi_{min}$ and for $\varphi_{wide}$-wide:

$$X_k = \left((L - z \tan\psi_{wide})\left(-½ + \frac{k-1}{N-1}\right), \frac{L}{2}\sin\frac{\varphi_{wide}}{2}\right), k = 1, \dots, N \quad (1)$$

In phase Quadrature (IQ) images were reconstructed on a polar grid (elevation $\varphi$, radius R), ($d\varphi$, $\lambda/2$) = (0.5°, 0.15 mm) from the RF data. The blood flow signal was retrieved from a block of frames using spatiotemporal filtering based on singular value decomposition (SVD) [8]. The filtered frames were squared and averaged into a final ultrasensitive power Doppler image.

Acoustic parameters were set to $N$ = 12 virtual sources of $\varphi_{wide}$ = 90° wide and $\psi_{min}$ = 10° giving, using (1), $X$ = ({-4.0, -3.3, -2.5, -1.8, -1.1, -0.4, 0.4, 1.1, 1.8, 2.5, 3.3, 4.0}, 9.1) mm at a 6-kHz pulse repetition frequency for $R_{im}$ = 100 mm deep. The wavelength resolution was $\lambda$=0.3 mm. A 1-s long UD acquisition at a 500-Hz frame rate, removing the first 10% eigenvalues from SVD, was set for Doppler imaging.

We used 89 slices with a step dθ = 2° for a full scan. Slice overlapping in elevation was estimated based on the mean piezo elements' height $h_{mean} = \frac{h}{\pi}\int_0^\pi \sin x\, dx$ = 6.4 mm with max height h = 10 mm, focal distance $f$ = 55 mm, and elevation at focus $el_f$ derived from diffraction $el_f = \frac{\lambda f}{h_{mean}}$ = 2.6 mm (Fig 1b).

For deep imaging, high power transmission is needed. Because of the small probe's aperture, the safety heating limit of +2°C for implantable device was surpassed. We thus implemented a pause of 11 s between each slice acquisition, limiting heating to +1.6°C. The total duration was 18 min, which remains acceptable for our application

*B. 3D Doppler scan processing*

3D reconstruction was performed by interpolating data on a cartesian grid. UD slices were first corrected with a depth attenuation compensation. Then we interpolated the data, from a polar slice with rotation, a spherical grid (azimuth θ, elevation φ, radius φ), (dθ, dφ, λ/2) = (2°, 0.5°, 0.15 mm) to cartesian grid (lateral x, elevation y, depth z), (0.25, 0.25, 0.25) mm³, using Delaunay triangulation (scatteredInterpolant, MatLab).

Relative 3D Doppler scan in percentage was computed from two successive 3D Doppler scans ($relativescan_{1/2} = \frac{scan_2 - scan_1}{scan_1} \times 100$). Scans were first registered with intensity-based rigid registration (imregtform, MatLab) using large vessels from thresholding. When registration was not possible, ROI comparisons were carried out and the non-uniform background amplitudes stemming from different depth of the probe were compensated.

Signal, image processing and 2D slice rendering were performed with MatLab (MathWorks, MA, USA). 3D rendering was performed with Amira software (Thermo Fisher Scientific Inc., MA, USA).

### III. IN VITRO FEASIBILITY

3D Doppler scan feasibility with our non-implantable preclinical prototype was first assessed in vitro on a phantom with tubes (MODEL ATS 524, CIRS, Norfolk, USA). The setup consisted of the prototype being placed above a water tank with the phantom immersed inside. The phantom's tubes were filled with blood mimicking fluid and connected to a pump to create flow at 4 cm/s. The phantom could be moved to image flow at different locations relative to the probe. 3D Doppler scan (18 min) was performed for each configuration (Fig 2a).

Then, the Contrast-to-Noise Ratio (CNR) of the Doppler signal in tubes was computed for all configurations. All CNR slices were averaged on a single slice, roughly fitting an 80°-wide sector field of view (Fig 2b).

### IV. IN VIVO PROTOCOL

In vivo imaging was carried out on six pigs with the collaboration of GIP Cyceron (Caen, France). The pig animal model was chosen because it offers a relatively large brain

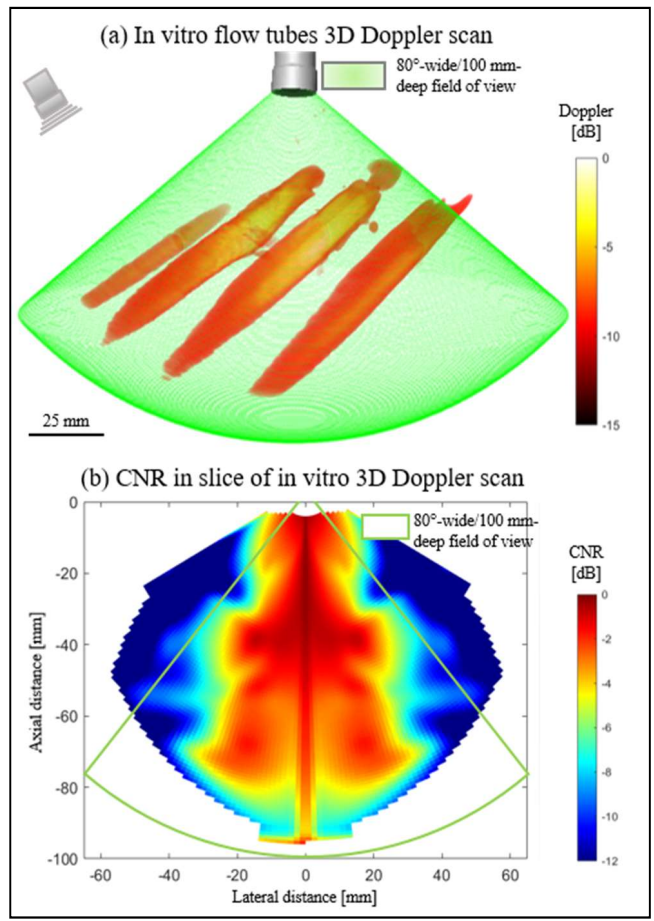

Fig 2. In vitro: (a) cartesian transformed 3D Doppler scan of four 8-mm diameter inclined tubes with flowing water with the 80°-wide and 100-mm deep conic field of view, (b) Contrast-to-Noise Ratio averaged in a slice from all tube's configurations.

(5 x 6 x 7 cm3) and is readily available. The animal was anesthetized and sedated, and a conventional 15-mm diameter burr hole was drilled through the skull at Bregma +20 mm and midline -12 mm up to the dura mater leaving brain protective layers intact. 2D real-time Bmode was displayed for positioning. Perfusion CT with iodinated contrast injection was also performed for stroke assessment with a CT scanner (Model, GE Healthcare, Chicago, USA).

Global blood flow variations were induced. CBV was decreased through mechanical hyperventilation inducing hypocapnia [9] and later increased through acetazolamide injection, a vasodilator [10]. A stroke model was carried out to induce ischemia. Iron chloride application on the carotid makes blood clots form in the vessel by oxidation. Then compressing the carotid, blood clots migrate into the brain vasculature [11]. A neurosurgeon assessed the ischemic core with perfusion CT [12] before and after iron chloride application.

### V. RESULTS

*A. Coverage*

First, the coverage of the brain through a burr hole was assessed. Using an anatomical CT scan, we registered a pig brain atlas (The Pig Imaging Group, University of Illinois Urbana-Champaign, USA) on the 3D Doppler scan. We measured about 60% brain coverage and up to 80% with good

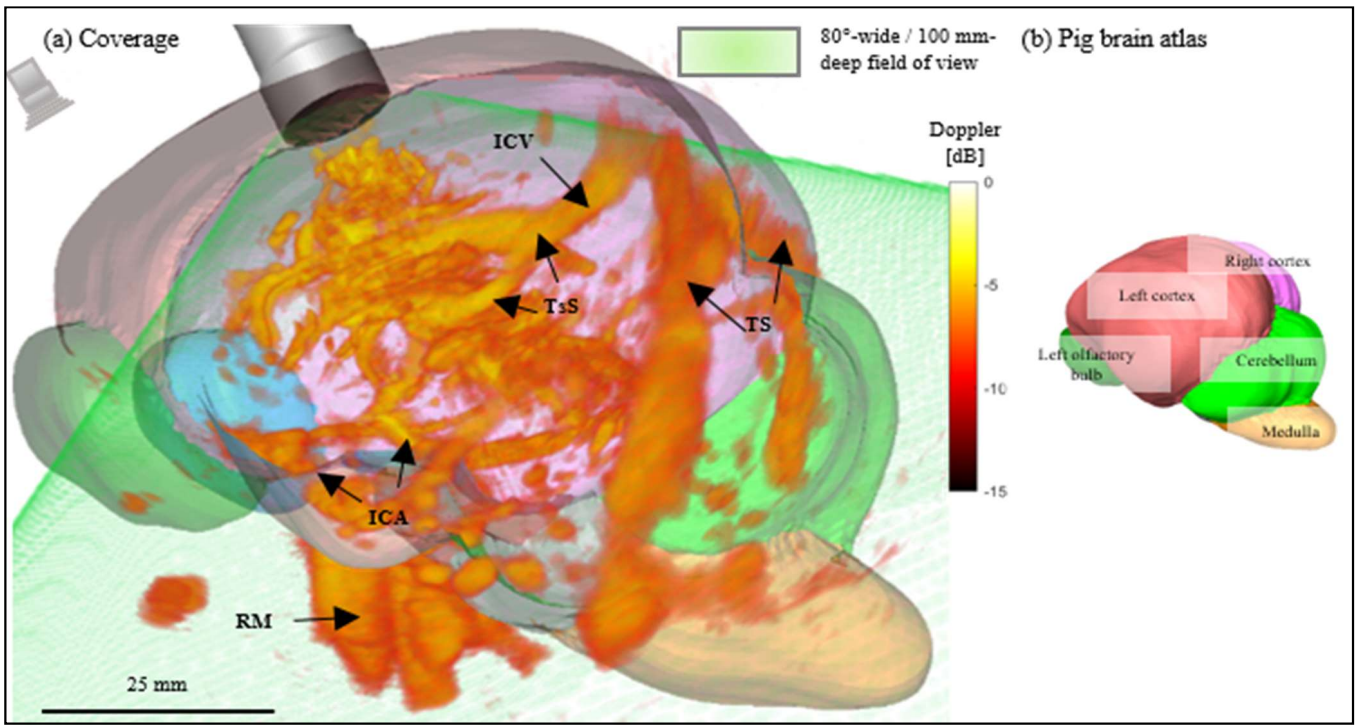

Fig 3. Pig brain coverage: (a) 3D Doppler scan of pig brain registered on pig brain atlas with the 80°-wide and 100-mm deep field of view with main vessels indicated ICV: internal cerebral vein, TsS: thalamostriate system, TS: transverse sinuses, ICA: internal carotid arteries and RM: rete mirabilis, (b) MR atlas.

probe positioning. The field of view of an 80°-wide cone was retrieved in vivo (Fig 3).

### B. Global cerebral blood volume variations

Then, global CBV variations were measured. First, hyperventilation induces a CBV decrease, and later, acetazolamide induces an increase. Two 3D Doppler scans were performed at baseline and when blood variations were established, allowing us to estimate relative 3D Doppler scans in percentage. In this example, we observed, as expected, a global decrease of 2.5% following mechanical hyperventilation (Fig. 4a) and a global increase of 6% following acetazolamide injection (Fig. 4b). The relative Doppler found in the whole brain was in accordance with induced CBV variations in 4/6 pigs for decrease and 4/5 pigs for increase

### C. Stroke

Finally, stroke detection was assessed following induced ischemia. For this pig, an ischemic stroke formed with perfusion dropping by 60% (14.0 to 4.9 a.u. CT's blood volume). The assessed ischemic core was found on both 3D Doppler scans performed before and after iron chloride application. In the core, fewer vessels were observed, and a decrease in Doppler signal of 25% was measured (0.161 to 0.105 normalized Doppler signal).

The formation of the ischemic core was explored with a higher frame rate. Right after iron chloride application, waiting for the ischemia to establish, a 2D Doppler movie of 60 frames in 10 min was acquired. In the assessed ischemic core, CBV was diminished by about 38%.

## VI. Discussion

The proof of concept of minimally-invasive whole-brain perfusion monitoring was assessed with a preclinical non-implantable prototype. A coverage of 60% of the brain was obtained, which allowed global blood flow monitoring and stroke detection.

A scan duration of 18 min was chosen to stay below the 2 degrees safety limitation for implantable device. The resulting scan duration is quite long for real-time imaging but remains acceptable in the context of perfusion and ischemia monitoring. Brain coverage was sufficient on the pig brain with 62 ± 12 % up to 80 % coverage. However, still missing superficial, extreme frontal, and occipital regions depending on the probe's relative orientation with respect to the brain. Nonetheless, the pig brain is still half the depth of the human one, and deep imaging over 40 mm could not be evaluated because of the skull's interface with the lower base.

Multiple 3D Doppler scans were performed to assess longitudinal CBV variations, close to the clinical application of monitoring. However, in our protocol, CT scans performed between the 3D Doppler scans required manually removing and replacing the probe, introducing shifts in the probe orientation and position, which would not have happened with an implantable device. If post hoc volume registration could handle most of those cases, this posed, in some cases, significant challenges for correctly assessing small CBV variations due to the non-uniform amplitude of the Doppler images due to the probe geometry.

The stroke monitoring could only be reliably performed in one over six pigs due to a combination of iron chloride model issues, probe repositioning issues, or premature animal death.

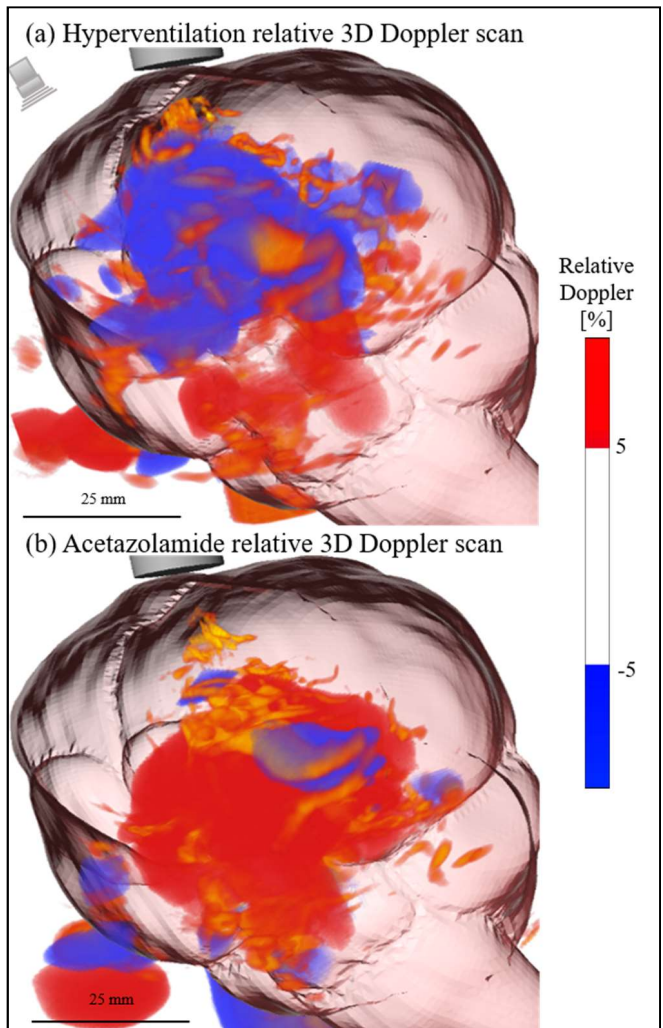

Fig 4, Global cerebral blood volume (CBV) variations: relative 3D Doppler scan of an induced CBV (a) decrease via mechanical hyperventilation, (b) increase via acetazolamide injection

For this case, the 2D Doppler movie clearly showed the ischemic core with a decrease of 38% in the ischemic core. Registered 3D Doppler scans with intensity normalization found a decrease of 25%.

In future experiments, we plan to couple Doppler with Bmode, elastography, and explore ULM scan without having to move the probe between the longitudinal scans.

## CONCLUSION

The proof of concept of minimally-invasive whole-brain brain perfusion monitoring was performed with a preclinical prototype on pigs. Scan duration and overall brain coverage allowed global perfusion monitoring and stroke detection. This device, which could be further miniaturized for implantation inside a burr-hole, could offer a promising brain perfusion monitoring tool for neurocritical care alongside ICP monitoring.

## ACKNOWLEDGMENT

Thanks to the Agence Nationale de la Recherche, which funded this project (Project ANR 20 CE19-0026). The GIP Cyceron team managed the experimental part with pigs, and Iconeus team advised us on the ultrasound sequence.

## REFERENCES

[1]  K. Asehnoune *et al.*, "The research agenda for trauma critical care," *Intensive Care Med*, vol. 43, no. 9, pp. 1340–1351, Sep. 2017, doi: 10.1007/s00134-017-4895-9.

[2]  R. L. Macdonald, "Delayed neurological deterioration after subarachnoid haemorrhage," *Nat Rev Neurol*, vol. 10, no. 1, pp. 44–58, Jan. 2014, doi: 10.1038/nrneurol.2013.246.

[3]  M. Wintermark *et al.*, "Comparative overview of brain perfusion imaging techniques," *Stroke*, vol. 36, no. 9, pp. e83-99, Sep. 2005, doi: 10.1161/01.STR.0000177884.72657.8b.

[4]  V. Hingot *et al.*, "Early Ultrafast Ultrasound Imaging of Cerebral Perfusion correlates with Ischemic Stroke outcomes and responses to treatment in Mice," *Theranostics*, vol. 10, no. 17, pp. 7480–7491, 2020, doi: 10.7150/thno.44233.

[5]  C. Demené *et al.*, "Multi-parametric functional ultrasound imaging of cerebral hemodynamics in a cardiopulmonary resuscitation model," *Sci Rep*, vol. 8, no. 1, Art. no. 1, Nov. 2018, doi: 10.1038/s41598-018-34307-9.

[6]  R. J. Forsyth, J. Raper, and E. Todhunter, "Routine intracranial pressure monitoring in acute coma," *Cochrane Database Syst Rev*, no. 11, p. CD002043, Nov. 2015, doi: 10.1002/14651858.CD002043.pub3.

[7]  C. Papadacci, M. Pernot, M. Couade, M. Fink, and M. Tanter, "High Contrast Ultrafast Imaging of the Human Heart," *IEEE Trans Ultrason Ferroelectr Freq Control*, vol. 61, no. 2, pp. 288–301, Feb. 2014, doi: 10.1109/TUFFC.2014.6722614.

[8]  C. Demené *et al.*, "Spatiotemporal Clutter Filtering of Ultrafast Ultrasound Data Highly Increases Doppler and fUltrasound Sensitivity," *IEEE Trans Med Imaging*, vol. 34, no. 11, pp. 2271–2285, Nov. 2015, doi: 10.1109/TMI.2015.2428634.

[9]  J. G. Laffey and B. P. Kavanagh, "Hypocapnia," *New England Journal of Medicine*, vol. 347, no. 1, pp. 43–53, Jul. 2002, doi: 10.1056/NEJMra012457.

[10] A. S. Vagal, J. L. Leach, M. Fernandez-Ulloa, and M. Zuccarello, "The acetazolamide challenge: techniques and applications in the evaluation of chronic cerebral ischemia," *AJNR Am J Neuroradiol*, vol. 30, no. 5, pp. 876–884, May 2009, doi: 10.3174/ajnr.A1538.

[11] T. Bonnard and C. E. Hagemeyer, "Ferric Chloride-induced Thrombosis Mouse Model on Carotid Artery and Mesentery Vessel," *JoVE (Journal of Visualized Experiments)*, no. 100, p. e52838, Jun. 2015, doi: 10.3791/52838.

[12] M. Wintermark, R. Sincic, D. Sridhar, and J. D. Chien, "Cerebral perfusion CT: technique and clinical applications," *J Neuroradiol*, vol. 35, no. 5, pp. 253–260, Dec. 2008, doi: 10.1016/j.neurad.2008.03.005.